\documentclass{nature}
\usepackage{graphicx, color, amsmath}
\usepackage[utf8]{inputenc}
\usepackage{amsmath}
\usepackage{color}
\usepackage{amsfonts}
\usepackage{amssymb}
\usepackage[dvipsnames]{xcolor}

\definecolor{red}{rgb}{1,0,0}
\definecolor{green}{rgb}{0,1,0}
\definecolor{orange}{rgb}{1,0.5,0}
\definecolor{blue}{rgb}{0,0,1}
\definecolor{purple}{rgb}{1,0,1}

\begin{document}

\title{Turbulent laser puffs }
\author{}


\author{Svetlana Slepneva\textsuperscript{1,2{*}}, Ben O'Shaughnessy\textsuperscript{1,2}, Andrei Vladimirov\textsuperscript{3,4}, Sergio Rica\textsuperscript{5,6}, and Guillaume Huyet\textsuperscript{7,8}}

\maketitle\\
$^1$Centre for Advanced Photonics and Process Analysis \& Department of Physical Sciences, Cork Institute of Technology, Cork, Ireland\\
$^2$ Tyndall National Institute, University College Cork, Lee Maltings, Dyke Parade, Cork, Ireland\\
$^3$ Weierstrass Institute, Mohrenstr 39, Berlin, Germany\\
$^4$Lobachevsky State University of Nizhny Novgorod, 23 Gagarina av., 603950 Russia\\
$^5$ Facultad de Ingenier\'ia y Ciencias, Universidad Adolfo Ib\'a\~nez, Santiago, Chile\\
$^6$ UAI Physics Center, Universidad Adolfo Ib\'a\~nez, Santiago, Chile.\\
$^7$ Universit\'e C\^ote d'Azur, CNRS, Institut de Physique de Nice, Nice, France\\
$^8$ National Research University of Information Technologies, Mechanics and Optics, 199034 Saint Petersburg, Russia\\
{*}corresponding author(s):
Svetlana Slepneva (svetlana.slepneva@tyndall.ie)

{\bf The destabilisation of laminar flows and the development of turbulence has remained a central problem in fluid dynamics\cite{JimenezScience,NatureComment} since Reynolds' studies in the 19th century\cite{Reynolds}. Turbulence is usually associated with complex fluid motions and most of the studies have so far been carried out using  liquids or gases. Nevertheless, on a theoretical viewpoint, turbulence may also arise in a wide range of fields such as biology and optics\cite{localized,turitsyna}. Here we report the results of experimental and theoretical investigation of the characteristic features of laminar-turbulent transition in a long laser commonly used as a light source in medical imaging and sensing applications\cite{canonical_oct,ssoct,OCTendosc}. This laminar to turbulence transition in the laser light is characterized by the appearance of turbulent puffs similar to those commonly observed in pipe flows and is accompanied by a loss of coherence and limits the range of applications\cite{FDMLpulses}. We present both experimental results and numerical simulations demonstrating that this transition is mediated by the appearance of a convective instability where localised structures develop into drifting bursts of turbulence, in complete analogy with spots, swirls and other structures in hydrodynamic turbulence\cite{avila,JimenezScience,NatureComment}. }

An everyday observation of spatio-temporal patterns created by the ascending smoke of a cigarette points us to the transition from the laminar flow near the tip of the cigarette to the turbulent motion upward. The laminar-turbulent transition\cite{Reynolds,avila} is a crucial problem of engineering and science motivated by many applications such as turbulent transport, atmospheric flows and airplane development while remaining a partially understood problem of fluid dynamics\cite{JimenezScience,NatureComment}. 
The transition from laminar to turbulent flow follows usually the same scenario, evolving from a stationary motion to well-defined oscillations eventually ending in a turbulent wake. 
The Poiseuille flow inside a pipeline demonstrates an interesting example of the transition to turbulence since the laminar flow is theoretically linearly stable for any speed\cite{drazin} (or Reynolds number) but experimental observations show the development of turbulence associated with the formation of a turbulent spot downstream indicating that a finite amplitude perturbation destabilises the laminar regime\cite{avila}.

The important role of nonlinearities was first understood phenomenologically in the 1950s after the pioneering work of L.~D.~Landau\cite{landau44}. The transition from a laminar to a turbulent state may be characterized by the notion of bifurcation which has been at the centre of important developments from the 60s up to the present\cite{CrossHohenberg}. Many features observed in a turbulent flow are closely connected to the formation of localised structures such as solitary waves or vortices.  These structures are naturally observed in nonlinear optics\cite{solitons,vortex}, which has become an excellent test-bed to investigate turbulence and the emergence of complexity in spatially extended systems\cite{kramer}. It was, for example, recently demonstrated that the transition to turbulence in a long cavity laser is associated with the clustering of dark and grey solitons\cite{turitsyna}.
 
Convective instabilities are commonly observed in open flow instabilities where a disturbance may grow in time as it travels away from the region of its birth so its amplitude grows at any point along the path before eventually decaying to its original value. This contrasts with the more conventional absolute instabilities where an initial disturbance spreads out along the entire unstable region\cite{landaukinetics,huerre,plaza,couairon}. In the convective regime, noise-induced spots, swirls and other structures are commonly observed in hydrodynamic turbulence. Noise may also lead to the appearance of a laminar-turbulent transition at finite distance in the convective regime as demonstrated by Deissler\cite{deisl}. Here we demonstrate that the light in a laser system with a very long cavity can exhibit various transitions from laminar to turbulent behaviour. In particular, we show that this transition may happen through the appearance of convective bursts of turbulence in the laser. 

\begin{figure}[h!]
a \includegraphics[width=10cm]{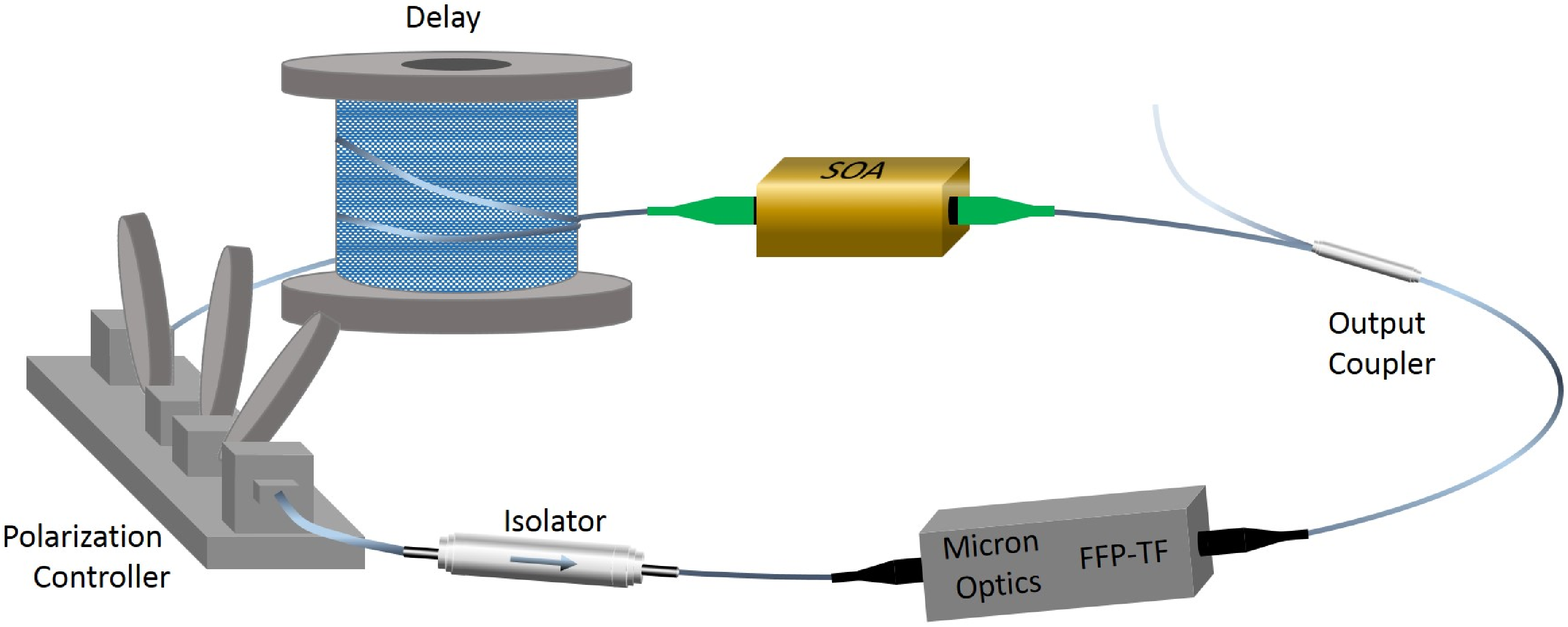}\\$\quad$ 
b \includegraphics[width=10cm]{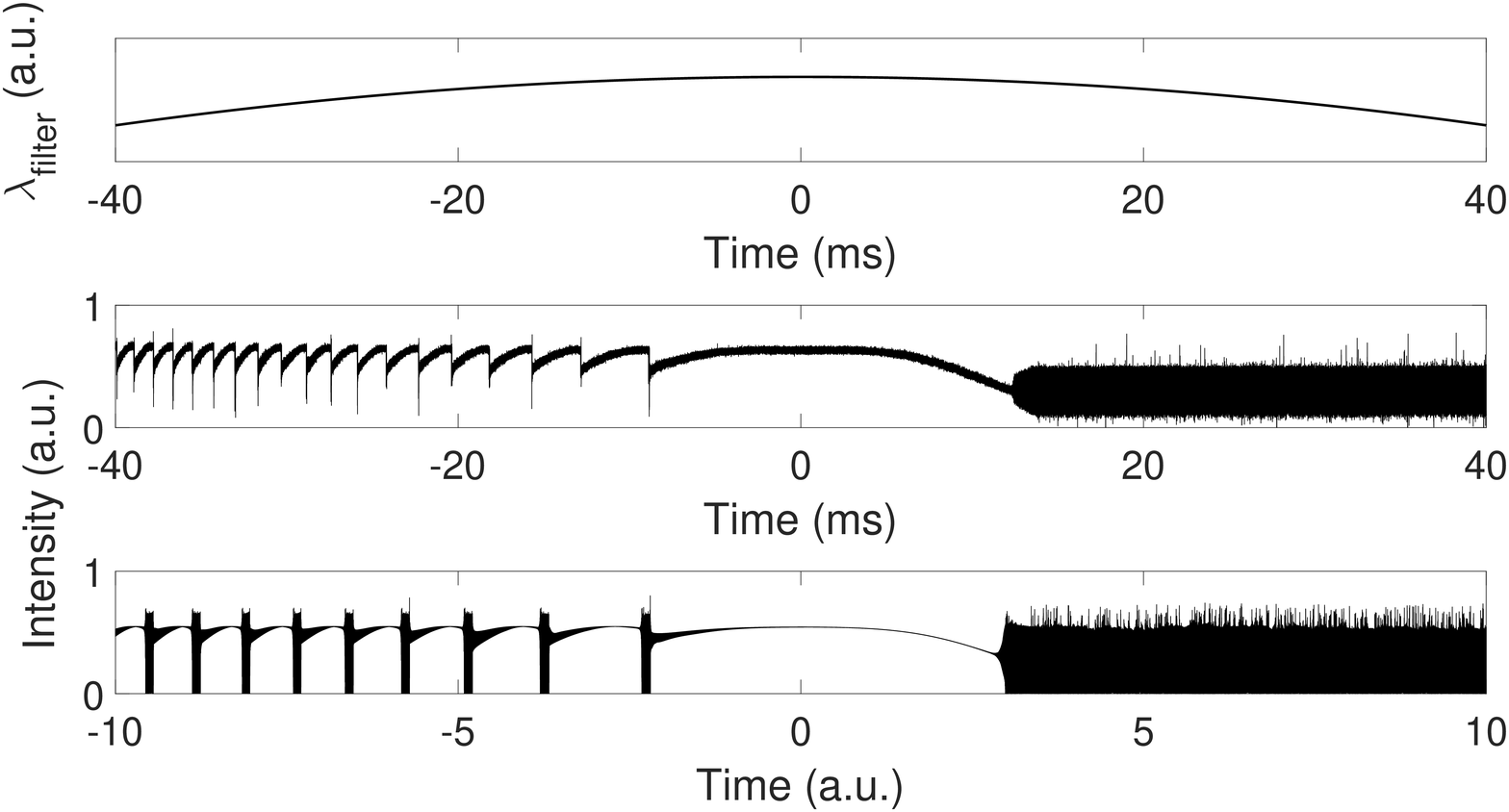}
\caption{\textbf{Subcritical and supercritical bifurcations in a wavelength swept laser.}  (\textbf{a}) Experimental set-up of a long cavity wavelength swept laser incorporating a semiconductor optical amplifier (SOA) and a fast Fabry-P\'erot tunable filter (FFP-TF) (See Methods for the details).  (\textbf{b}) The filter wavelength variation in time (top). Experimental (middle) and theoretical (bottom) observations of subcritical and supercritical modulational instabilities. A sequence of subcritical bifurcations occurs as the filter transmission wavelength increases while a single supercritical bifurcation leads to the appearance of a turbulent state as the the filter transmission wavelength decreases. The experimental time trace was recorded for a laser with a 17m cavity length and 1Hz filter modulation frequency. The theoretical time trace was obtained by direct numerical simulation of Eq.~(\ref{eq:E}) and ~(\ref{eq:G}). \newline}
\label{fig:s}
\end{figure}

We investigate the wavelength swept laser 
experimentally (Fig.~\ref{fig:s}a) and theoretically using the following 
delayed differential equations that describe the temporal evolution of the laser gain $G(t)$ and the electric field envelope $A(t)$. These equations read\cite{slepneva13}
\begin{eqnarray}
\dot A(t)+A(t)-i\Delta\left(t\right)A(t) &= &\sqrt{\kappa}e^{\left(1-i\alpha \right)G\left(t-T\right)/2}A\left(t-T\right),\label{eq:E}\\
\dot G(t) &=&\gamma\left[g_{0}-G(t)-\left(e^{G(t)}-1\right)\left|A(t)\right|^{2}\right],\label{eq:G}
\end{eqnarray}
where $\Delta(t)$ is the central frequency of the sweeping filter, $T$ is the roundtrip time of light in the cavity, $g_0$, $\gamma$, $\alpha$ and $\kappa$ are the unsaturated gain, normalized gain relaxation rate, linewidth enhancement factor and loss per cavity roundtrip, respectively. 

If the filter is moved quasi-statically, we observe both experimentally and theoretically, that the single mode laser operation undergoes either a subcritical or a supercritical Hopf bifurcation depending on the direction of the filter tuning. The increase of the filter transmission wavelength results in a subcritical bifurcation that ultimately leads to the emergence of another single mode red detuned solution. The decrease of the filter transmission wavelength results in a supercritical bifurcation which leads to the emergence of a turbulent regime. Such behaviour, as illustrated in Fig.~\ref{fig:s}b, can be observed when the filter wavelength is periodically tuned at a very low frequency. 

A similar situation occurs when the filter is tuned with a period $T_F$ close to the cavity roundtrip time $T$. This regime is known as Fourier Domain Mode-Locking (FDML)\cite{FDML} since the optical frequency of the laser follows the periodic evolution of the central frequency of the filter $\Delta (t)$. The interest in FDML lasers is mostly driven by its application in swept source Optical coherence tomography (OCT)\cite{canonical_oct,ssoct,OCTendosc} and more recently, Raman spectroscopy\cite{FDMLraman}. This laser, due to its unprecedented high sweep rates, enabled volumetric video-rate real time OCT imaging whose quality relies on ultra-fast broadband swept sources with a long coherence length. However the loss of coherence within the sweep leads to deterioration of the image quality and limits the generation of extremely short pulses to 60ps after optical decompression\cite{FDMLpulses}. 
  
\begin{figure}[h!]
a \includegraphics[width=6cm]{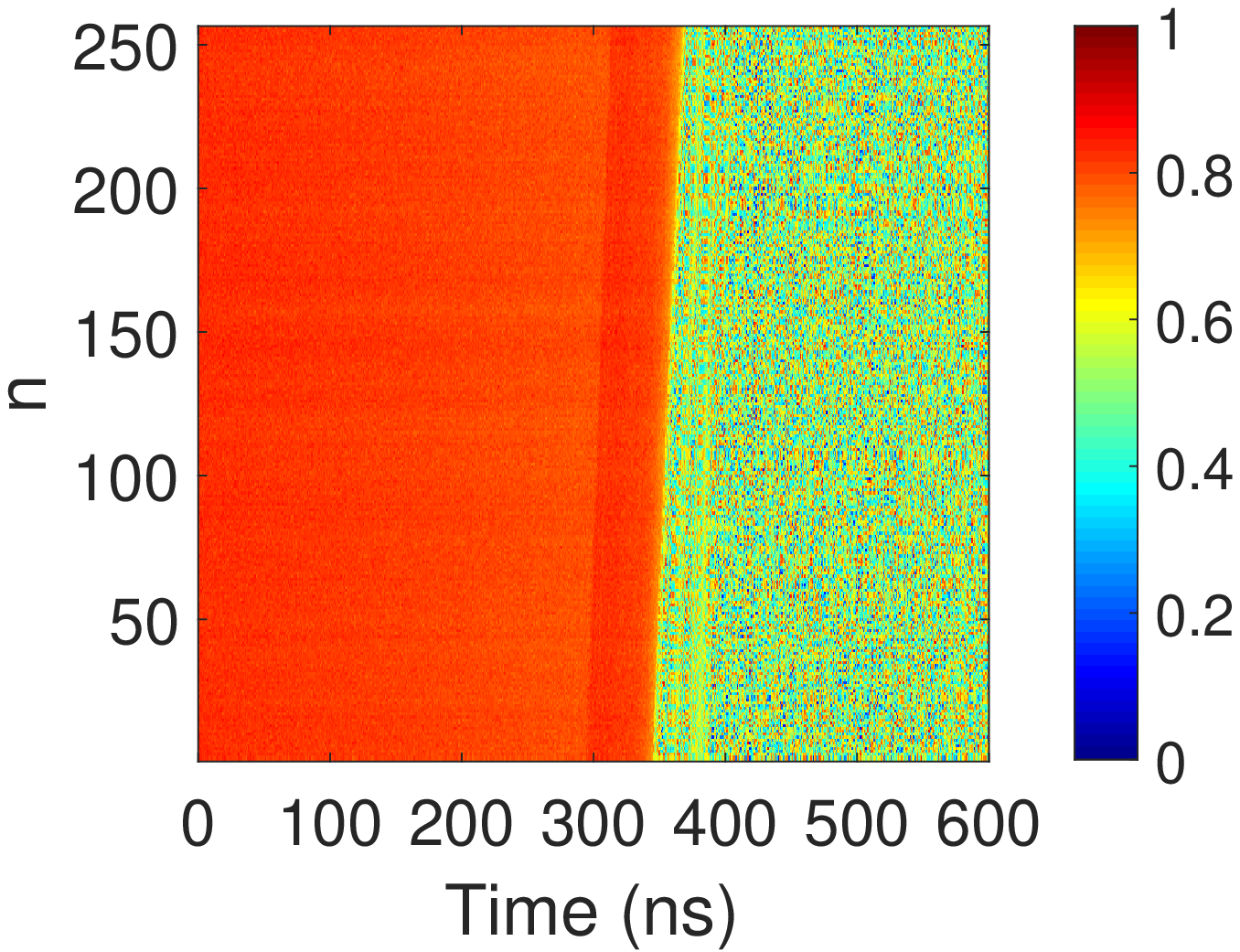}
c  \includegraphics[width=6cm]{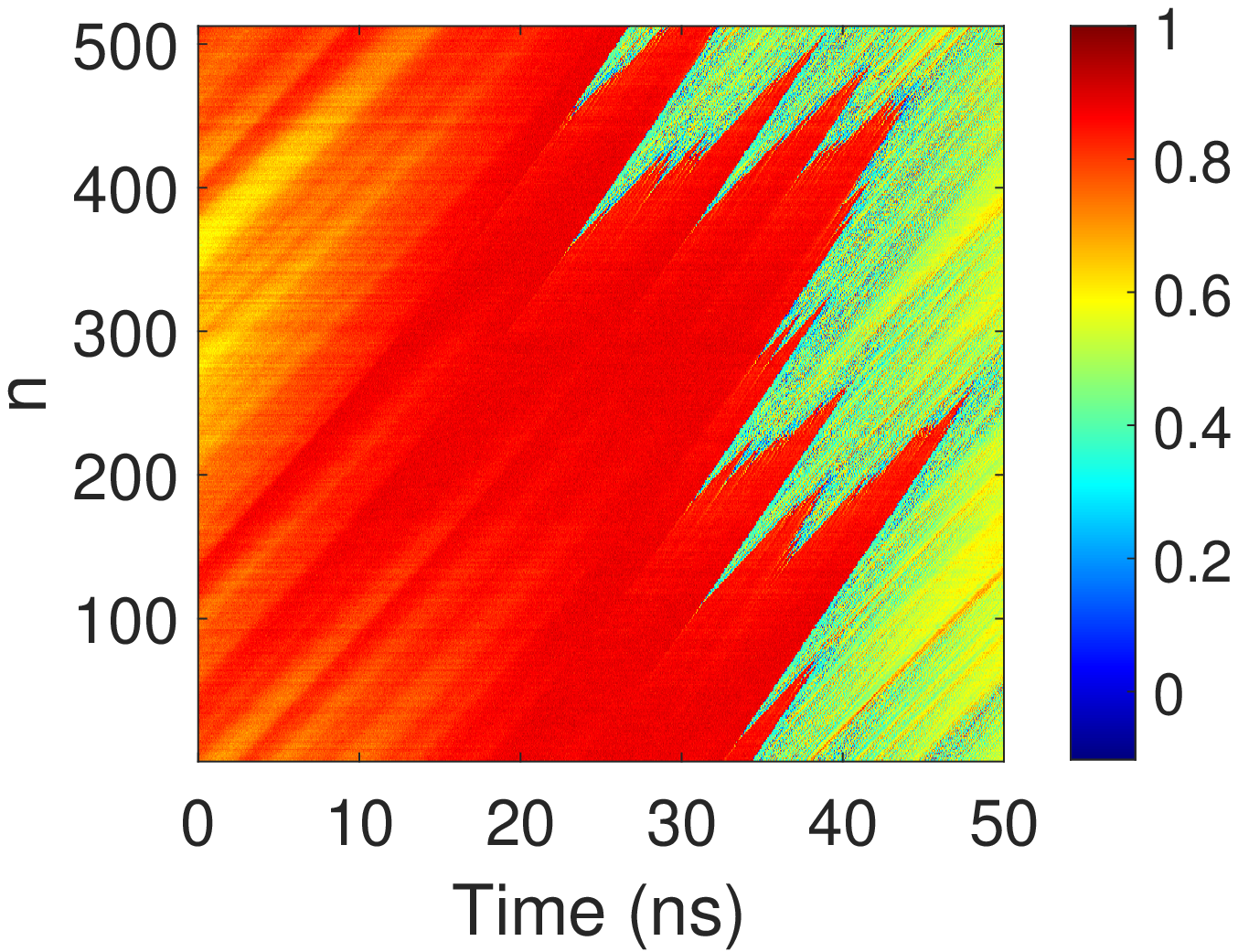}\\
b  \includegraphics[width=6cm]{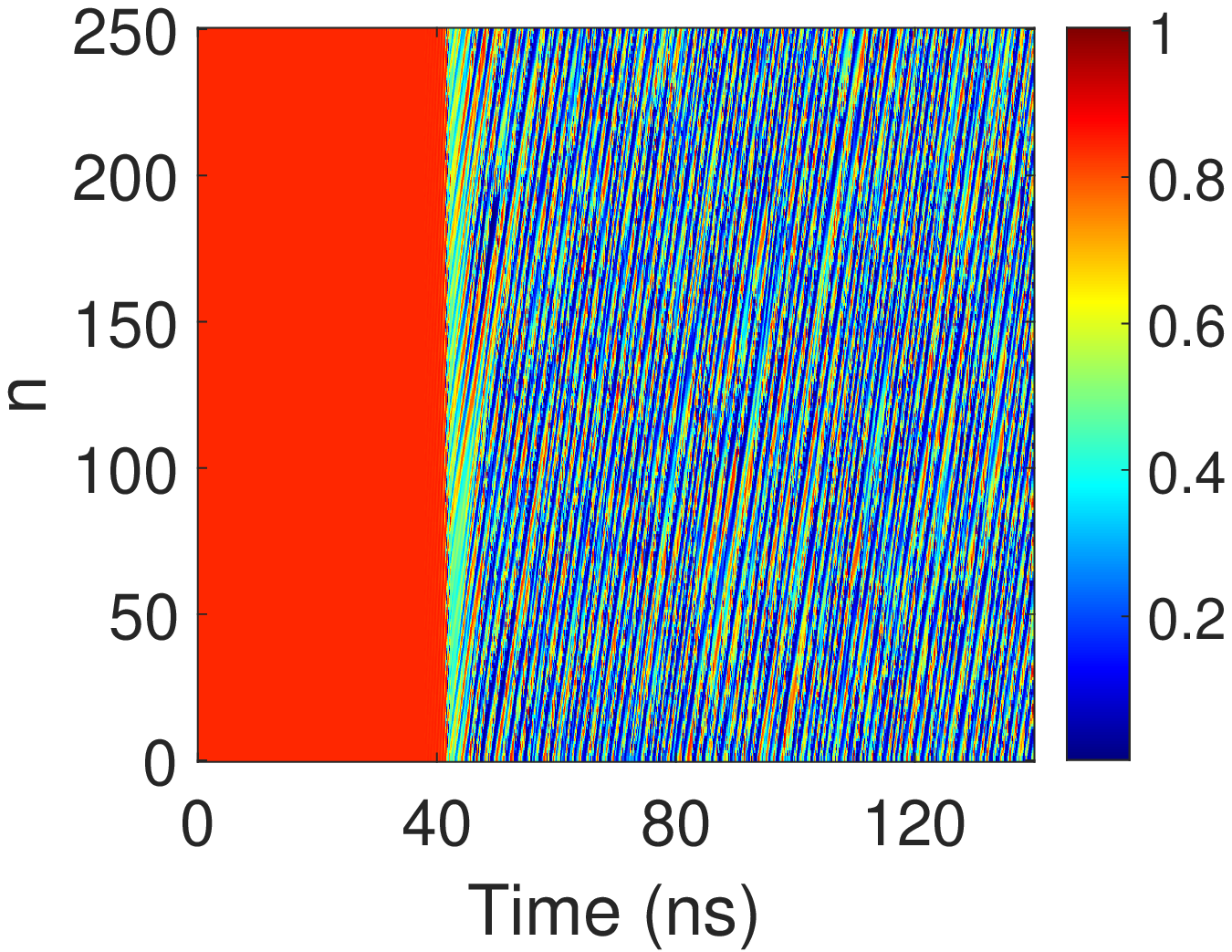}
d  \includegraphics[width=6cm]{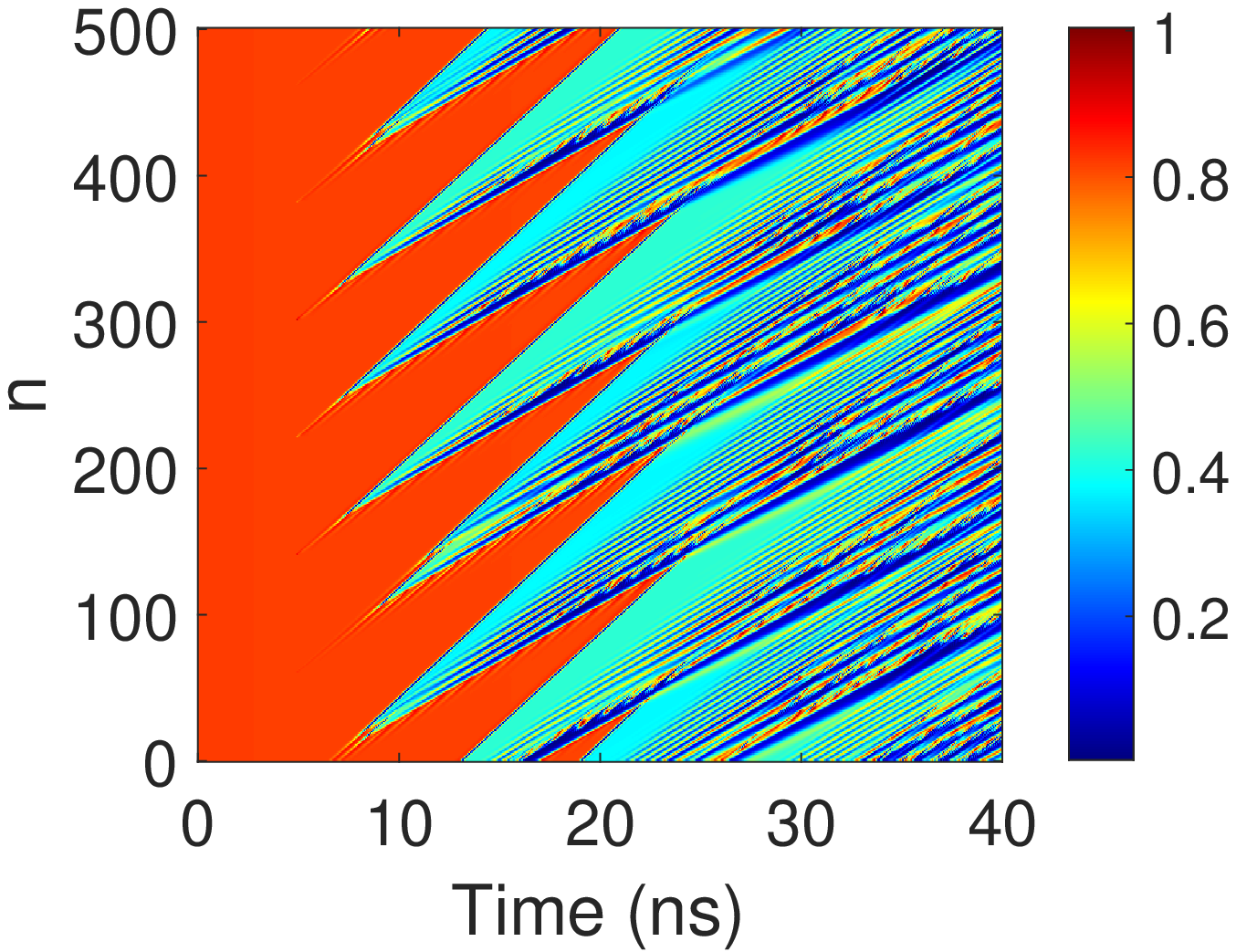}
\caption {\textbf{Evolution of the laser intensity near the transition between the laminar and turbulent regimes for subsequent filter periods (n).} Experimental  \textbf{(a)} and theoretical \textbf{(b)} 2D diagrams (see Methods) of the laser intensity evolution in the absolute instability regime corresponding to $\delta_{c_1}<\delta <\delta_{c_2}$. Experimental \textbf{(c)} and theoretical \textbf{(d)} 2D diagrams of the laser intensity evolution in the convective instability regime corresponding to $\delta_{c_2}<\delta <\delta_{c_3}$. The triangular features in \textbf{(c)} and \textbf{(d)} represent the emergence of localised structures from the laminar regime. These structures drift toward the turbulent region with a speed $v\sim 1/p_{up}$ and the size of the turbulent spot grows at a rate  $r\sim 1/p_{low}-1/p_{up}$, where $p_{up}$ and $p_{low}$ are the slopes of the upper and lower boundaries between the laminar state and the localised structures. The colorbar represents the normalised laser intensity. \newline
}
\label{fig:2D}
\end{figure}

\begin{figure}[h!]
a   \includegraphics[width=6cm]{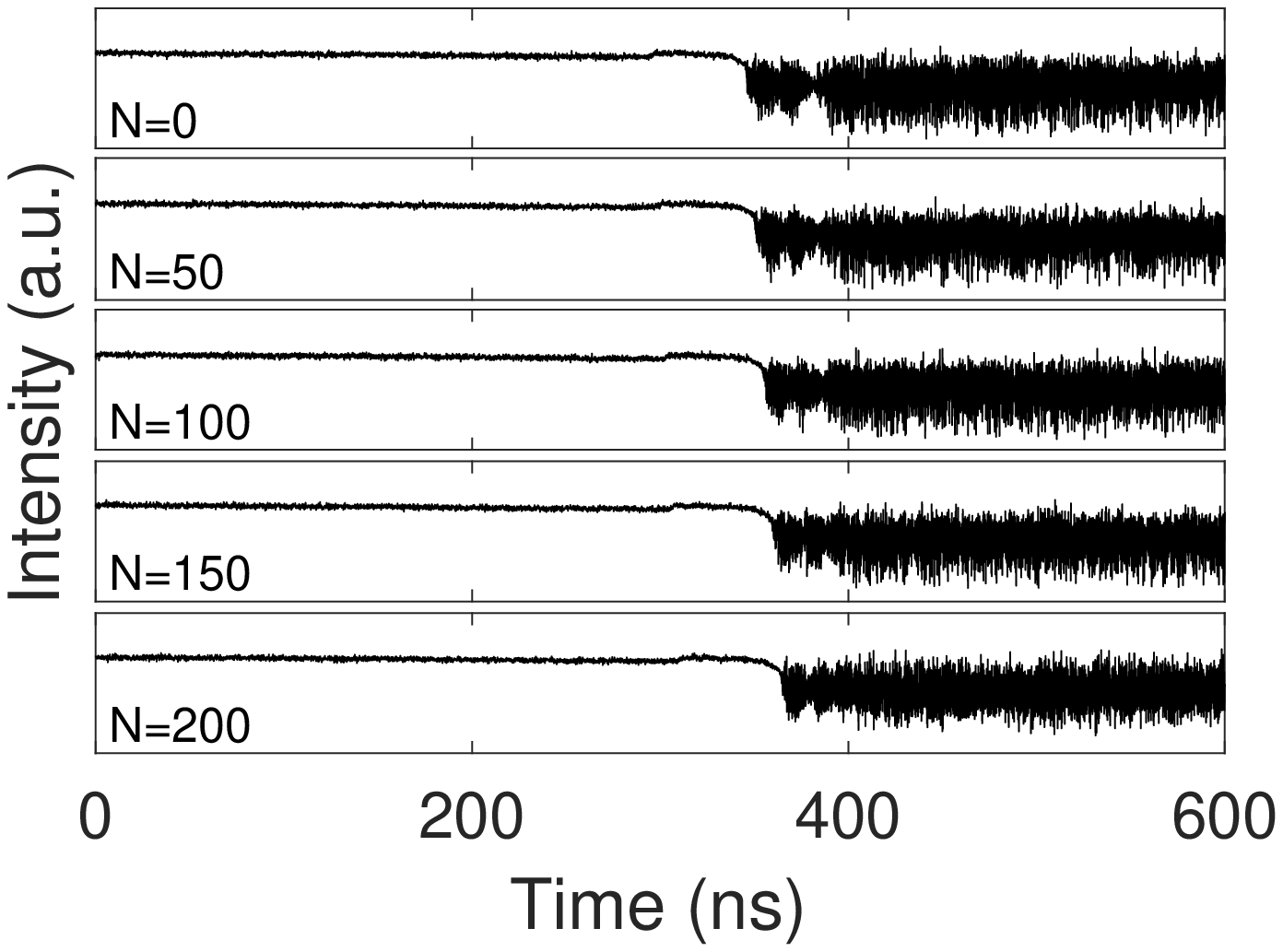}
c    \includegraphics[width=6cm]{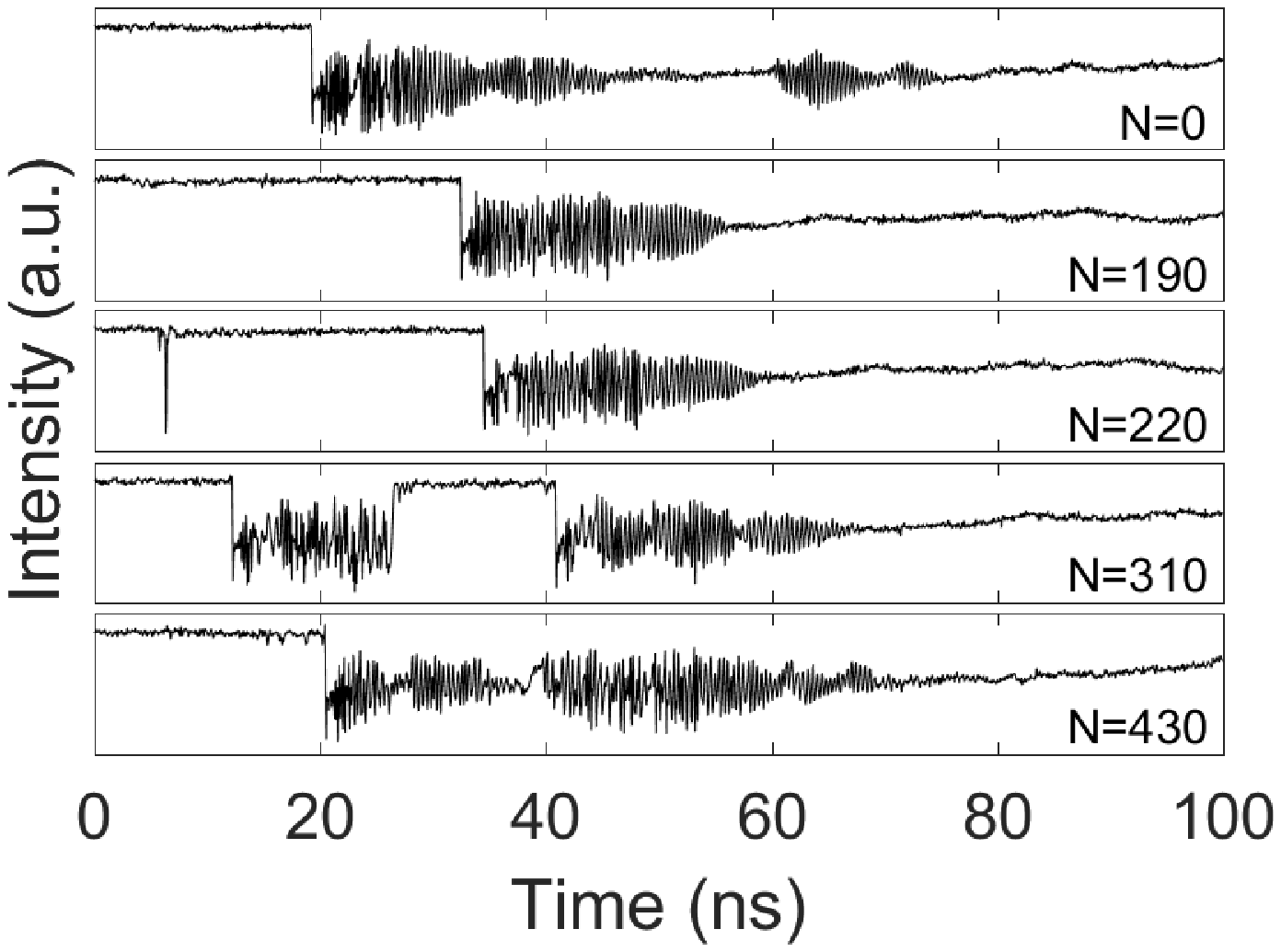}\\
b  \includegraphics[width=6cm]{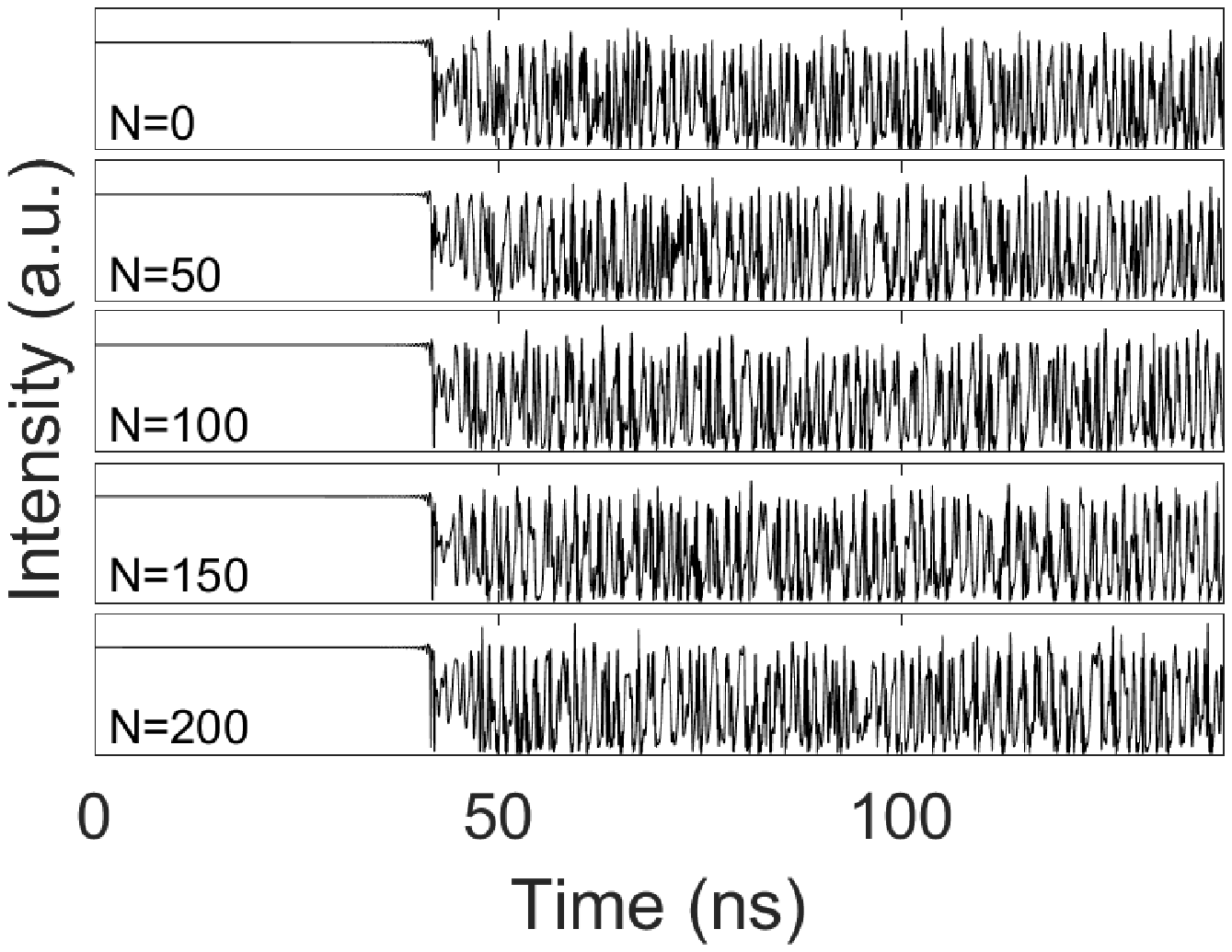}
d  \includegraphics[width=6cm]{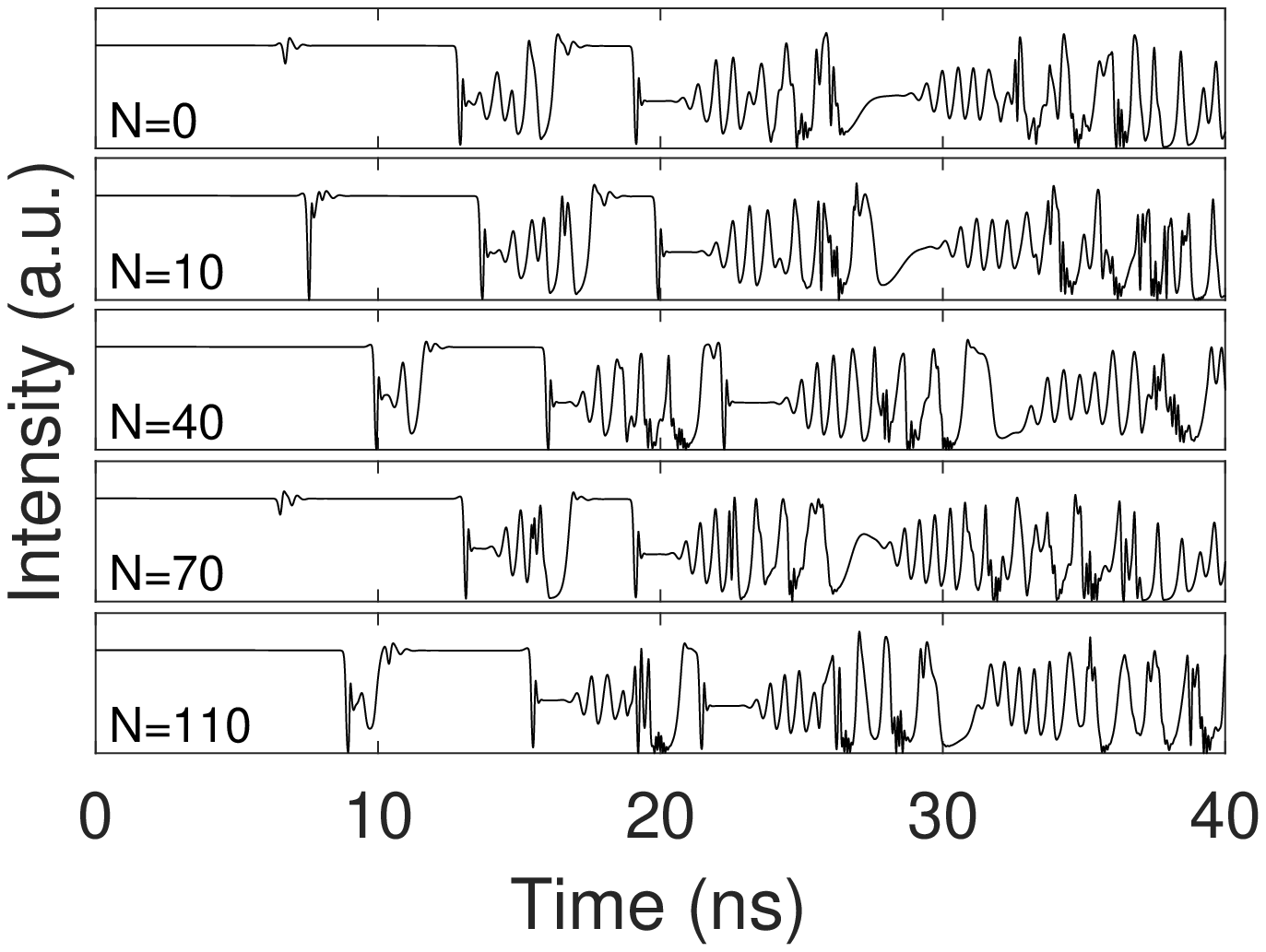}
\caption{ 
\textbf{ Detailed analysis of the laser intensity evolution near the subcritical bifurcation point.} The laser intensity was recorded at $t=nT_F$, where $n$ is the number of the filter cycles of modulation. Experimental \textbf{(a)} and theoretical \textbf{(b)} observations of the laser intensity demonstrating a sharp transition from laminar to turbulent regimes at a small value of detuning $\delta_{c_1}<\delta <\delta_{c_2}$. This corresponds to the absolute instability regime. Experimental \textbf{(c)} and theoretical \textbf{(d)} observations  of creation and drift of turbulent fronts in the convective regime ($\delta_{c_2}<\delta <\delta_{c_3}$). The nucleation of a hole structure that appears suddenly in the cavity is shown in \textbf{(c)} at round trip $n=220$.
Full evolution of the laminar to turbulent transition observed experimentally and theoretically for the both absolute and convective regimes are shown in Supplementary Movies. \newline
}
\label{fig:manyplots}
\end{figure}

When the filter is driven in a perfect resonance with the cavity roundtrip time, the laser emits a set of frequency modulated solutions  corresponding to single mode solutions in the filter frame\cite{slepneva13}. These single mode solutions have the same stability criteria as the single mode solutions observed when the central frequency of the filter is fixed. If the filter is tuned out of resonance with the cavity round trip, the stability of these frequency modulated solutions depends on detuning $\delta = 2\pi (1/T_F-1/T)$. While the behaviour of the supercritical bifurcation remains qualitatively the same for any values of $\delta$, the dynamics around the subcritical point strongly depends on its value and, to investigate this point further, we consider the dynamics for the increasing values of $\delta >0$.  As $\delta$ increases, we identify four different regimes. For $\delta<\delta_{c_1}$, the laminar regime remains stable for the entire sweep. For $\delta_{c_1}<\delta <\delta_{c_2}$, the laminar regime becomes turbulent as soon as it passes the bifurcation point; the transition between laminar and turbulent regimes remains at a fixed value of the filter frequency. This regime is consistent with an absolute instability as described in Fig.~\ref{fig:2D}a-b and Fig.~\ref{fig:manyplots}a-b. Note, that the numerical simulation, integrated in the limit of a very large delay near the bifurcation point (see Methods), has demonstrated close correspondence with the experimental results. 
\newpage

\begin{figure}[h!]
a \includegraphics[width=8cm]{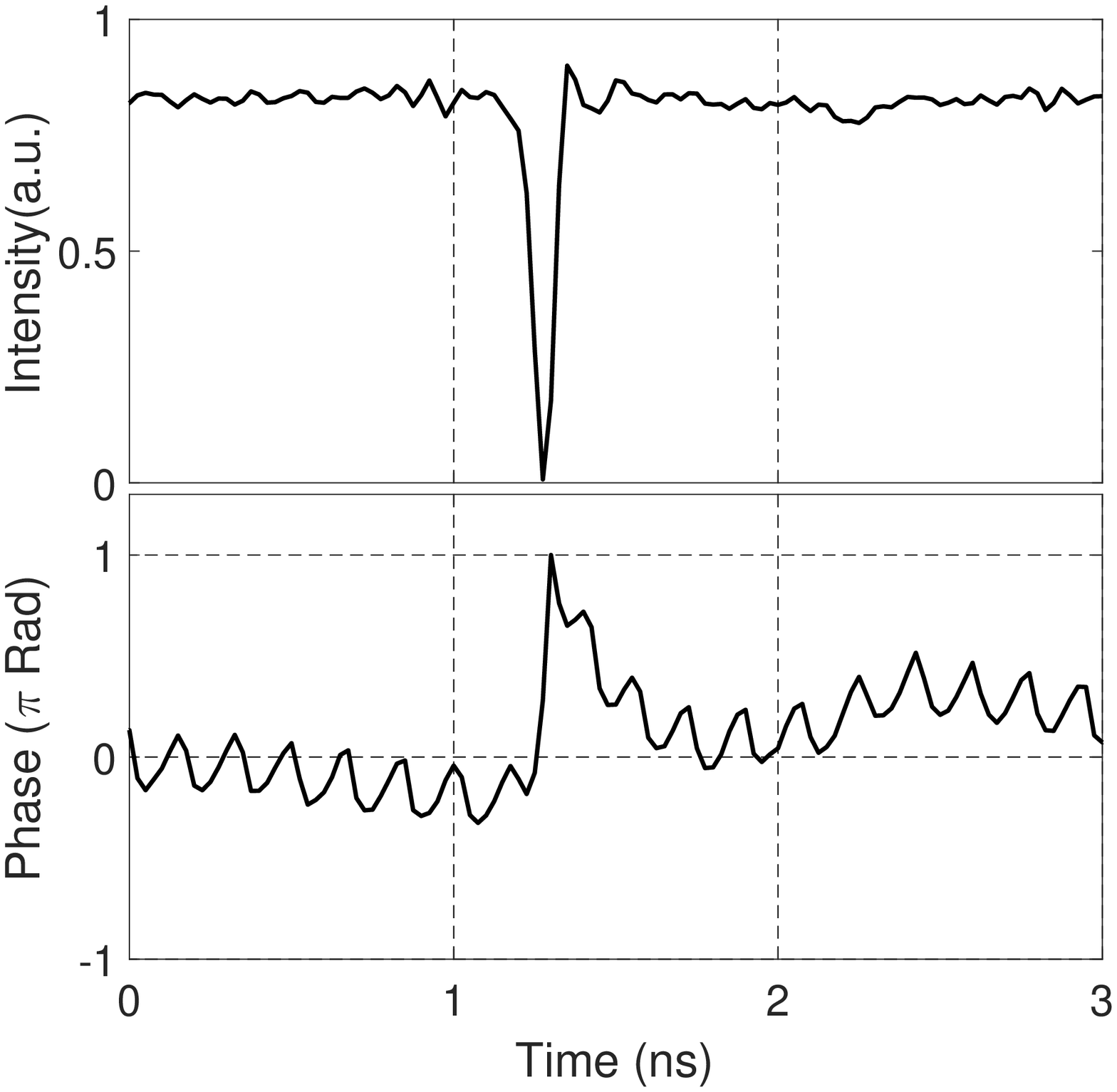}
b \includegraphics[width=8cm]{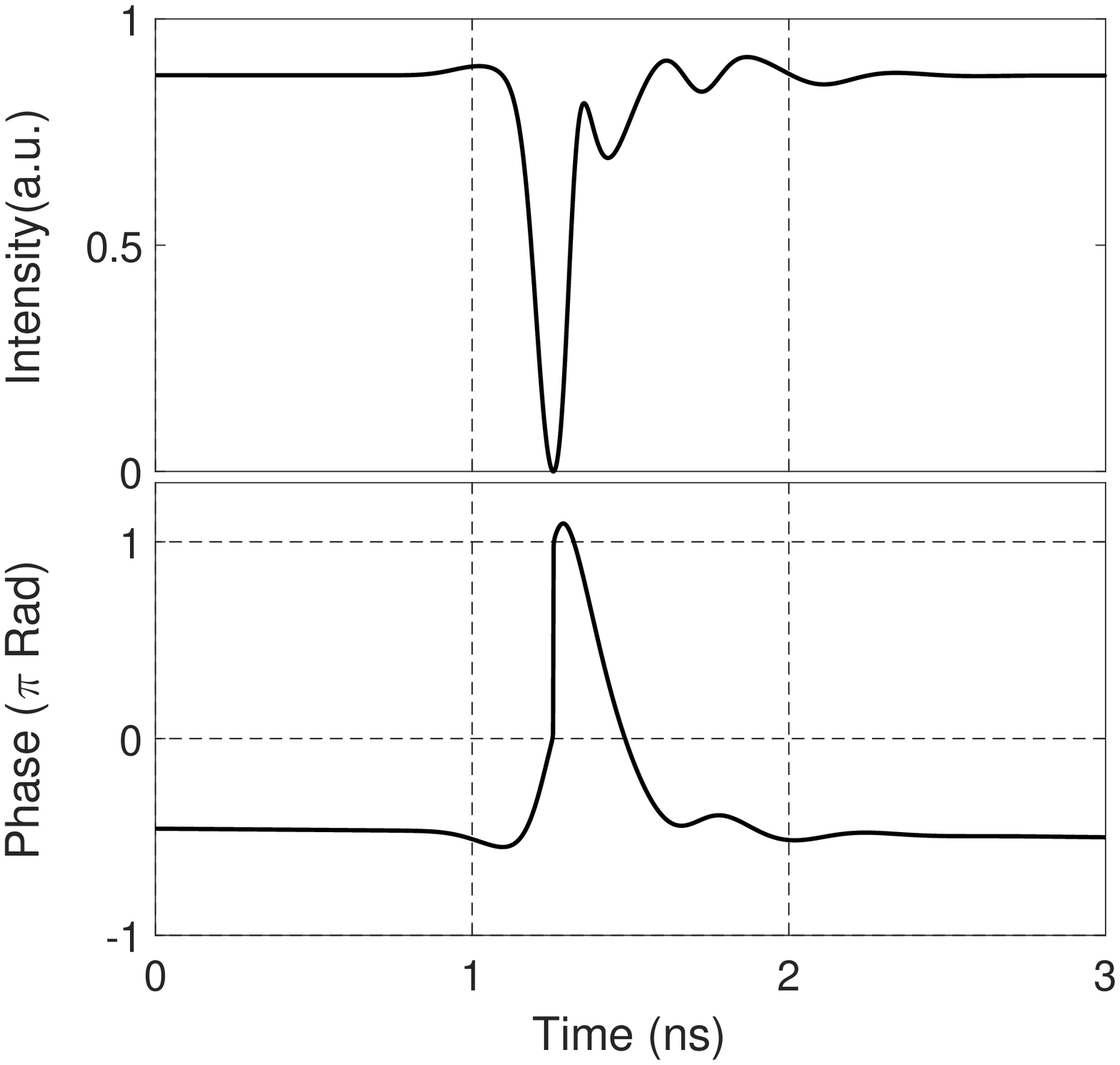}
\caption{\textbf{A single Nozaki-Bekki hole.} Experimentally observed \textbf{(a)} and theoretically modelled \textbf{(b)} intensity (top) and phase (bottom) of a  single Nozaki-Bekki hole. }
\label{fig:pulse}
\end{figure}

\begin{figure}
a \includegraphics[width=8cm]{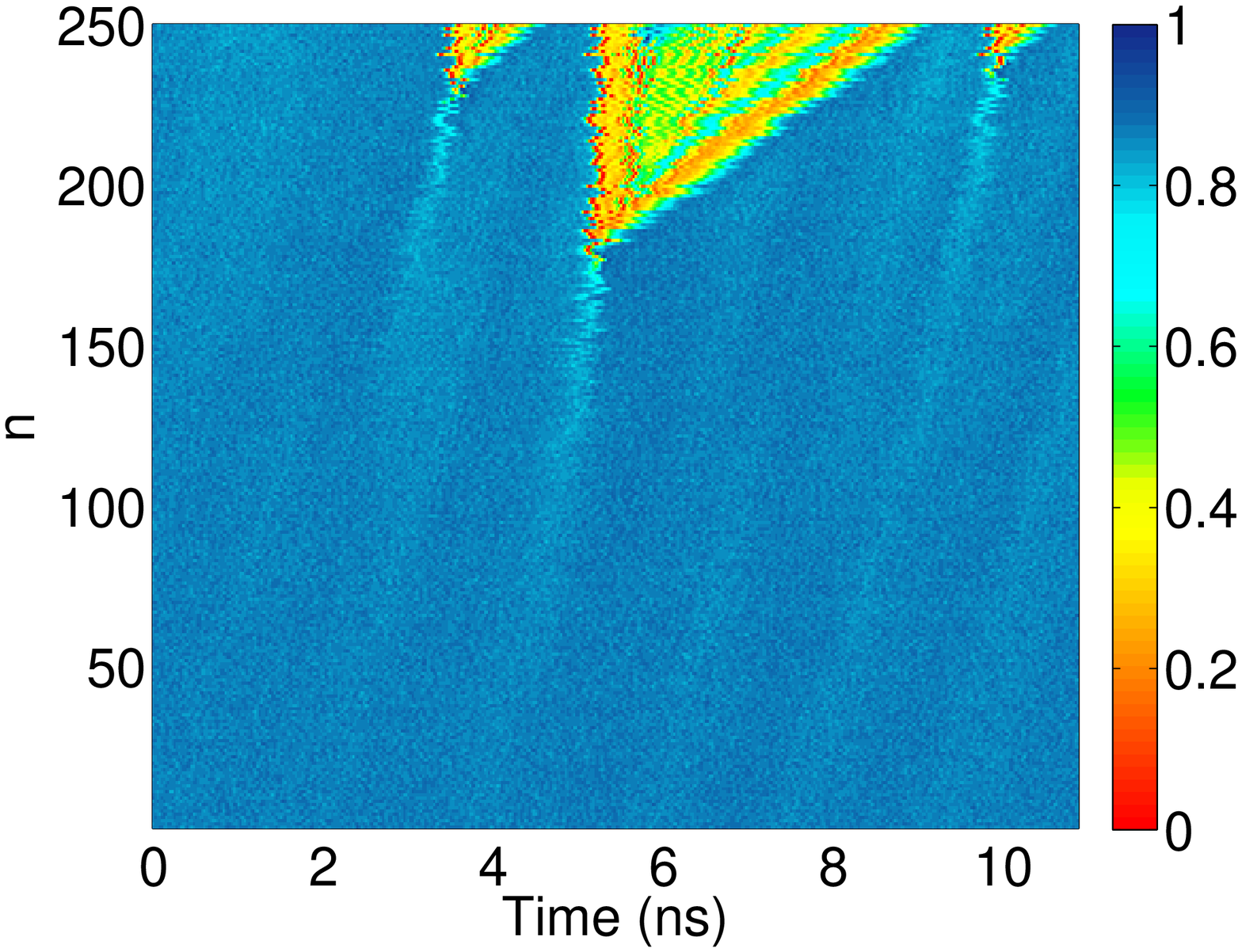}
b \includegraphics[width=8cm]{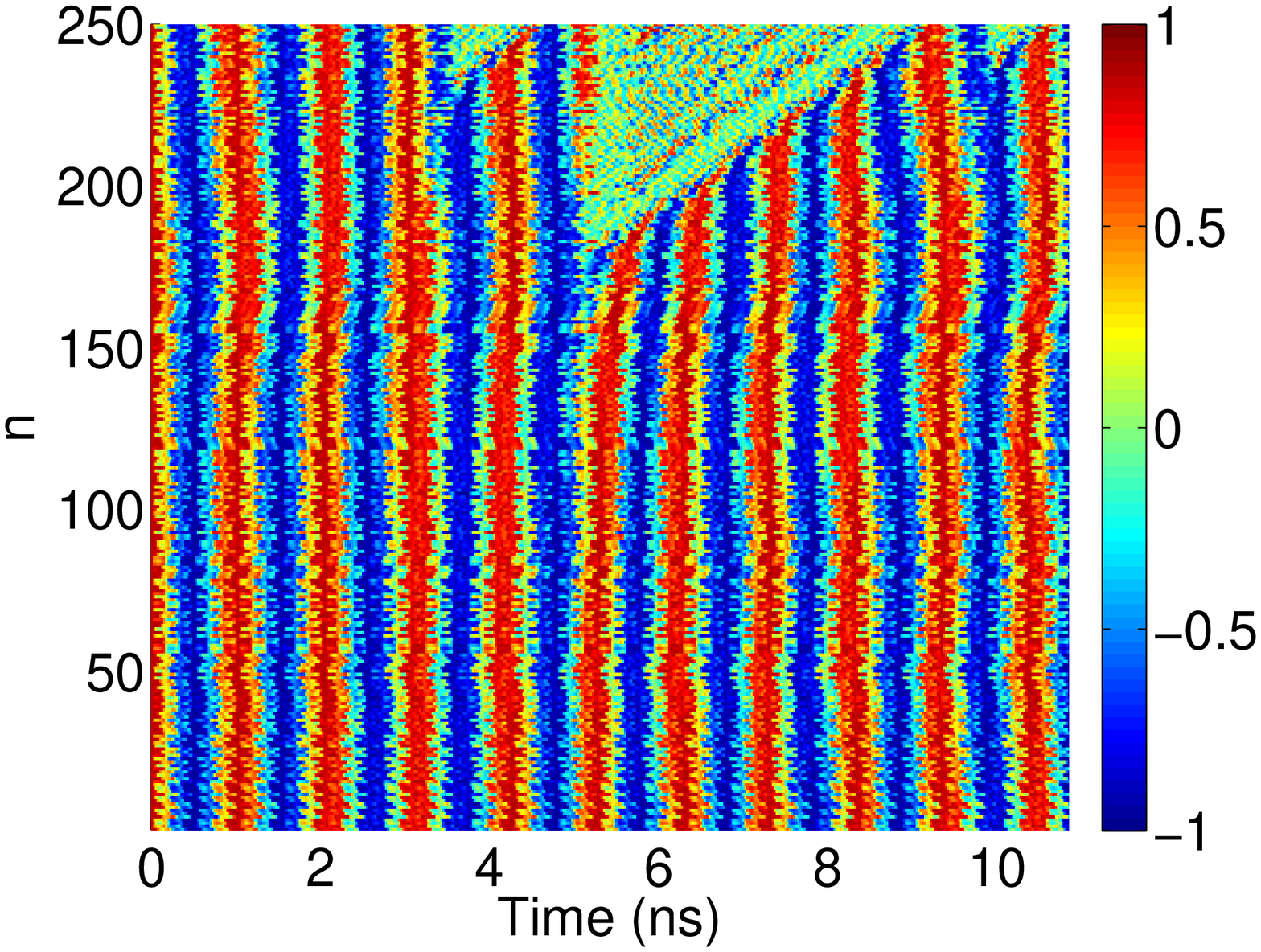}
\caption{ \textbf{Creation of three neighbouring transients. }
\textbf{a}, experimentally measured temporal evolution of the laser intensity  for 250 subsequent filter periods showing the emerging of three Nozaki-Bekki holes that initiate three turbulent regions within the laminar regime.  
\textbf{b}, experimentally measured real part of the associated electric field.}
\label{fig:re}
\end{figure}

For the larger detuning values, \textit{i.e.} $\delta_{c_2}<\delta< \delta_{c_3}$, the laminar regime can invade the turbulent regime at a rate much faster than the growth rate of the turbulent regime (Fig.~\ref{fig:2D}c-d, Fig.~\ref{fig:manyplots}c-d). Here the front between the laminar and turbulent regimes is no longer synchronised with the filter position but returns at every roundtrip time. In the filter frame this appears as a front drifting with a ``speed'' proportional to the detuning. As the laminar regime invades the turbulent region, we observe the spontaneous noise induced creation of bursts of turbulence that are mediated by the formation of spatio-temporal topological defects. For $\delta_{c_3}<\delta$, the laser displays a turbulent regime throughout the entire sweep. Experimentally, the values for $\delta_{c_1}$, $\delta_{c_2}$  are on the order of a few mHz while  $\delta_{c_3}$ is on the order of a few hundreds of mHz.
The temporal evolution of the laser intensity near the subcritical bifurcation point as a function of the filter period number $n$ is also shown in Fig.~\ref{fig:manyplots}. Fig.~\ref{fig:manyplots}a-b demonstrate the sharp transition 
and Fig.~\ref{fig:manyplots}c-d depict the regime where convective instability occurs displaying the drift of the laminar regime in the turbulent regime and the creation of localised structures. For example, Fig.~\ref{fig:manyplots}c shows the experimentally observed formation of a localised structure near $t=10ns$ for $n=220$. 
A measurement of the temporal evolution of the laser intensity and phase with a $25ps$ temporal resolution shows that the laser intensity reaches zero with an associated $\pi$-phase shift (Fig.~\ref{fig:pulse}a). This phase measurement technique (see Method) was also used to analyse the temporal evolution of the optical field. 
The 2D diagrams of the laser intensity (Fig.~\ref{fig:re}a) and the associated real part of the electric field (Fig.~\ref{fig:re}b) show the initial stage of creation of three localised structures.
Numerical simulations also show the existence of localized structures with the instantaneously vanishing electric field (Fig.~\ref{fig:pulse}b). This hole solution, known as Nozaki-Bekki hole\cite{Nozaki1,Nozaki2}, propagates ``down-stream'' with the group velocity (Fig.~\ref{fig:manyplots}c, $n=310$ and $n=430$). The Nozaki-Bekki hole nucleates the meta-stable turbulent phase that propagates at a larger speed ``downstream'', as demonstrated experimentally and numerically in Fig.~\ref{fig:manyplots}c-d. These turbulent regions grow while drifting and merging with other turbulent regions. As a result, the laminar region continues to invade the turbulent region and other turbulent domains are created.

\newpage
The hole solutions observed both experimentally and numerically can be associated to the family of Nozaki-Bekki holes observed in complex Ginzburg-Landau equation\cite{Nozaki1,Nozaki2}. These holes are asymmetric wave sources, emitting different wave numbers up and down streams\cite{vanSaarlos}, and it has been suggested that they are the building-blocks of spatio-temporal chaos in convection experiments\cite{Burguete}. It is, to our knowledge, the first experimental observation of holes inducing the turbulent transition in optical systems. 

We have demonstrated, experimentally and numerically, a scenario for the laminar-turbulent transition in a very long cavity wavelength swept laser. By precisely controlling the detuning between the cavity roundtrip time and the sweep period we are able to generate the transition between the absolute instability to the route to turbulence via creation of convective Nozaki-Bekki holes. This optical analogue of turbulent spots in the open flow in hydrodynamics, well-known as Poiseuille flow, is the mechanism that leads to a deterioration of the coherence of the laser. We believe that this detailed experimental study, coupled with our simple theoretical investigation will encourage further investigation and development of such lasers as well as expand their application.

\newpage
{\bf Methods}

{\bf Experimental setup.}
The FDML laser design included a 20km long cavity, corresponding to a 100 microsecond round-trip time, incorporating a semiconductor amplifier and a narrow bandwidth (80pm) Fabry-P\'erot tunable filter. The wavelength was sweeping over several nm around 1300nm central wavelength. The temporal evolution of the output power was measured by coupling the light to a high speed detector and analysed on a 12GHz real time oscilloscope. To measure the phase, the laser output was coupled to an unbalanced Mach-Zehnder interferometer those output was analysed through a $3\times 3$  interferometer in order to capture the phase difference $\phi(t)-\phi(t-T)=\eta(t)$ between the two arms of the interferometer\cite{linewidth}. Since this phase difference corresponds to the integral of the frequency over a time $T$, $\eta(t)$ is directly proportional to the (coarsed-grained) optical frequency averaged over the time $T$.

Experimental 2D diagrams describing the evolution of the laser intensity after several hundreds of successive modulation periods of the filter were obtained by measuring the laser intensity during some small time window of length $T_W$, repeated with a period matching the filter sweep period $T_F$. $T_W$ is much shorter than the cavity roundtrip time $T_R$. For each filter sweep number $n$, we examine the electric field of the laser in the interval $nT_F<t<nT_F+T_W$. The electric field is then treated as a function of two variables: time and filter period number $n$. A 2D picture of the laser dynamics is thus created for a small part of the filter sweep. The theoretical simulations follow a similar approach. Such diagrams are commonly used to draw the analogy between delay dynamical systems and one-dimensional spatio-temporal systems\cite{Giacomelli-PRL96}.

{\bf Numerical simulations.} 
The laser used for experiments has a long cavity (20km). The delay differential equation model could be used to run simulations of this laser but the delay would be very large, requiring lots of computation. The dynamics of interest are in a small time window  where laminar flow drifts into a region of instability and becomes turbulent over many sweep periods. The simplest case is close to the filter turning point, where the filter moves slowly and the laminar flow is a single mode solution to the DDE model, instead of a chirped FDML solution. By using the analytical solution for the laminar part and by assuming the cavity roundtrip time to be arbitrarily long, we can simulate only the required portion of each filter sweep (Tab.~\ref{tab:parameters}).

The electric field during the $n$th sweep is $A_n (t) = A(t+n T_F)$. We consider the case where $T=T_F+\tau$ so that the electric field drifts in the positive $t$ direction relative to the filter sweep profile with each subsequent period. In order to simulate $A_n (t)$ for $0<t<T_W$ we need $A_{n-1}(t)$ for $-\tau<t<T_W-\tau$. For $t>0$ this history comes from initial conditions and previous simulation. We assume that the filter is static at zero frequency for an arbitrarily long time before $t=0$ so that a single mode solution can be calculated analytically in this region, filling in the history for $-\tau<t<0$.

\begin{figure}
\includegraphics[width=\textwidth]{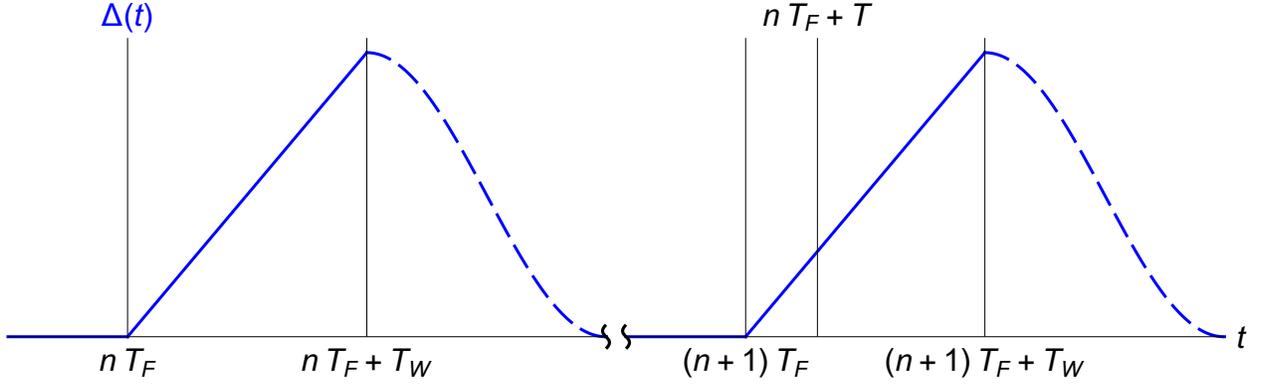}
\caption{Plot of filter sweep profile, $\Delta(t)$, used for simulations. For each sweep period $n$ the model equations are solved analytically for $t \leq n T_F$ and numerically integrated for $n T_F < t < n T_F+T_W$. The dashed region $t>n T_F+T_W$ is ignored.} 
\label{fig:simexplain}
\end{figure}

The full filter sweep used for simulation is as follows. The filter starts at frequency $f_0$ at $t=0$ and decreases linearly during the simulated time interval. After the simulated interval the filter returns to $f_0$ and remains there for the remainder of the arbitrarily long filter sweep, allowing the system to return to single mode operation before the beginning of the next simulated interval. The filter sweep profile of the real laser follows a sine wave. Near the turning point the filter is almost static for a time which is long compared with the time scale of the laser dynamics.

There is some equivalence between the single mode operation and chirped FDML operation. The following transformation puts the system in a frame where the filter is always at zero frequency, $A(t)=a(t) \exp \left( i\int_{t-T_R}^{t} \Delta(t')dt' \right)$. If the filter period is perfectly tuned to match the cavity round trip time then this system is identical to a system with a static filter, but the single mode solutions in this 'filter frame' system correspond to chirped FDML solutions. In the case of small detuning these solutions drift away from the filter over many round trips. Laminar FDML thus has similar instabilities to single mode operation, so a study of the single mode case provides insight into the transition to turbulence in FDML.

\begin{table}[h!]
\centering
\caption{Parameter values for simulations}
\medskip
 \begin{tabular}{|c|c|c|} 
 \hline
 Parameter & Description & Value \\ 
 \hline\hline
 $\gamma$ & Normalized carrier relaxation rate & 0.14  \\
 \hline
 $\kappa$ & Linear attenuation factor & 0.04  \\
 \hline
 $\alpha$ & Linewidth enhancement factor & 5.0  \\
 \hline
 $g_0$ & Unsaturated gain per roundtrip & 5.0 \\
 \hline
 \end{tabular}
 \label{tab:parameters}
\end{table}

\section*{Supplementary material}

Supplementary Movie SM1 visualises the experimental evolution of the laser intensity in the absolute instability regime, corresponding to a small value of detuning and demonstrates a sharp transition from laminar to turbulent regimes.  Supplementary Movie SM2 shows theoretical modelling of the laser intensity in the absolute regime.
Supplementary Movie SM3 visualises the experimental observation of laser intensity corresponding to the convective regime observed at a high detuning value. The movie shows the transition between the cw and turbulent regimes and the appearance of dropouts in the cw regime, illustrating the abrupt drop of the laser intensity to zero. These drops are the Nozaki-Bekki holes which evolve and drift towards the turbulent regime. Supplementary Movie SM4 shows theoretical modelling of the laser intensity evolution in convective regime. 

\section*{Acknowledgements}
The authors thank Stephen Hegarty for help with the experiment, Andrei Vladimirov, Evgeny A. Viktorov and Natalia Rebrova for useful advice and encouraging discussions. 
G.H. acknowledges the Science Foundation Ireland under Contract No. 11/PI/1152. S.S. gratefully acknowledges the
support of the EU FP7 Marie Curie Action FP7-PEOPLE-
2010-ITN through the PROPHET project, Grant No. 264687.

\section*{Author Contributions}
G.H. conceived the idea and initiated the study. S.S. implemented the experiment and performed the measurements. A.V. developed and analysed the model, B.O. ran the simulations. G.H., S.S., A.V., S.R. analysed and interpreted the experimental results. All the authors wrote the paper.

\section*{Competing Financial Interests statement}
The authors declare no competing financial interests.

\end{document}